\documentclass[twocolumn,showpacs,amsmath,amssymb,superscriptaddress]{revtex4}
\usepackage{graphicx, dcolumn, dsfont}
\begin{document}

\title{Efficient quantum circuits for arbitrary sparse unitaries}

\author{Stephen P. Jordan}
\affiliation{Institute for Quantum Information, Caltech, Pasadena, CA
  91125, USA}
\author{Pawel Wocjan}
\affiliation{School of Electrical Engineering and Computer Science,
  University of Central Florida, Orlando, FL 32816, USA}

\date{\today}

\newcommand{\ud}{\mathrm{d}}
\newcommand{\braket}[2]{\langle #1|#2\rangle}
\newcommand{\bra}[1]{\langle #1|}
\newcommand{\ket}[1]{|#1\rangle}
\newcommand{\Bra}[1]{\left<#1\right|}
\newcommand{\Ket}[1]{\left|#1\right>}
\newcommand{\Braket}[2]{\left< #1 \right| #2 \right>}
\renewcommand{\th}{^\mathrm{th}}
\newcommand{\tr}{\mathrm{Tr}}
\newcommand{\id}{\mathds{1}}

\begin{abstract}
Arbitrary exponentially large unitaries cannot be implemented
efficiently by quantum circuits. However, we show that quantum
circuits can efficiently implement any unitary provided it has at most
polynomially many nonzero entries in any row or column, and these
entries are efficiently computable. One can formulate a model of
computation based on the composition of sparse unitaries which
includes the quantum Turing machine model, the quantum circuit model,
anyonic models, permutational quantum computation, and discrete time
quantum walks as special cases. Thus we obtain a simple unified proof
that these models are all contained in BQP. Furthermore our general
method for implementing sparse unitaries simplifies several existing
quantum algorithms.
\end{abstract}

\maketitle

\section{Quantum Algorithm}

Following \cite{Aharonov_Tashma} we call an $N \times N$ matrix
$V$ combinatorially block diagonal if there exists a permutation
matrix $P$ such that $PVP^{-1}$ is block diagonal and the largest
block is $d \times d$ for $d = \mathrm{polylog}(N)$. We say $V$ is
row-computable if there is a polynomial-time algorithm which, given a row
index $i \in \{1,2,\ldots,N\}$, computes the values of all of the
nonzero matrix elements in row $i$ of $V$. In \cite{Aharonov_Tashma}
Aharonov and Ta-Shma show that quantum circuits can efficiently implement all
row-computable combinatorially block diagonal unitaries. More
precisely, one can implement $U$ satisfying $\|U-V\| \leq \epsilon$
using $\mathrm{poly}(\log N, d, \log(1/\epsilon))$ quantum gates. As a
corollary to this, quantum circuits can also efficiently simulate time
evolution according to combinatorially block diagonal
Hamiltonians. That is, one can implement $U$ satisfying $\|U-e^{-i H
  t}\| \leq \epsilon$ using $\mathrm{poly}(\log N, d,
\log(1/\epsilon))$ gates\cite{Aharonov_Tashma}. 

We call a matrix row-sparse if each row has at most polynomially many
nonzero entries. A row-sparse Hamiltonian can always be written as a
sum of polynomially many combinatorially block diagonal
Hamiltonians. Thus, using the Trotter formula, one can implement the
time evolution $e^{-i H t}$ for any $N \times N$ $d$-sparse Hamiltonian
$H$ to precision $\epsilon$ using a quantum circuit of
$\mathrm{poly}(\log N, d, \|Ht\|, 1/\epsilon)$
gates\cite{Aharonov_Tashma}. Subsequent work has 
improved the efficiency of such Hamiltonian
simulations\cite{Childs_thesis, Cleve_sim}. This suggests the question
of whether sparse unitaries can generically be implemented by quantum
circuits of polynomially many gates. As shown below, the answer is
yes.

Let $U$ be an $N \times N$ unitary such that each row or column has at
most $d$ nonzero entries, and these entries are efficiently
computable. Adapting ideas
from \cite{Harrow} we let
\begin{equation}
\label{hu}
H = \left[ \begin{array}{cc} 0      & U \\
                             U^\dag & 0
\end{array} \right].
\end{equation}
It is easy to see that $H$ is a row-sparse, row-computable Hermitian
matrix, $H^2 = \id$, and $\|H\| = 1$. Because $H^2 = \id$ it follows
that
\begin{equation}
\label{sincos}
e^{- i H \theta} = \cos(\theta) \id - i \sin(\theta) H.
\end{equation}
As discussed above, the time evolution $e^{-iHt}$ induced by any
row-sparse row-computable $N \times N$ Hamiltonian $H$ can be
implemented to precision $\epsilon$ by a quantum circuit of
$\mathrm{poly}(\log N, d, \|Ht\|, 1/\epsilon)$ gates. By equation 2,
choosing $t = \pi/2$ yields 
\begin{equation}
e^{-iHt} = -i \left[ \begin{array}{cc} 0      & U \\
                                      U^\dag & 0
\end{array} \right].
\end{equation}
Thus,
\begin{equation}
\label{final}
e^{-iH\pi/2} \ket{1} \ket{\psi} = -i \ket{0} U \ket{\psi}.
\end{equation}
The global phase $-i$ is irrelevant. The action $\ket{1} \to \ket{0}$
on the ancilla qubit can be made into the identity by adding an initial
NOT gate.

This technique works for all sparse efficiently computable unitaries,
but it is not restricted to sparse unitaries. It works for any $U$
such that $H$, as given by equation \ref{hu}, can be efficiently
simulated. Recent results of Childs\cite{Childs_nonsparse} show, among
other things, how to efficiently simulate any Hamiltonian $H$ whose
entrywise absolute value has at most polynomially large operator norm.
By our results, this implies efficient quantum circuits for all
unitaries satisfying the same condition. Furthermore, any future
advances in Hamiltonian simulation will imply, via the technique given
here, a corresponding advancement in efficient quantum circuit
implementation of unitaries.

Simple counting arguments show that arbitrary (non-sparse) unitaries
on $n$ qubits require exponentially many quantum gates to
implement\cite{Nielsen_Chuang}. A simple procedure using $O(2^{2n})$
quantum gates to implement an arbitrary unitary is given in section
4.5 of \cite{Nielsen_Chuang}. Far more sophisticated techniques have
subsequently been devised to construct more efficient (but still
exponentially large) quantum circuits implementing arbitrary unitaries
on $n$ qubits\cite{Shende, Brennan, Tucci, Nakajima1, Nakajima2, Berry}.

\section{Applications}

We can formulate a model of quantum computation based on sparse
unitary matrices. On $n$ qubits, we consider any row-sparse,
column-sparse, row-computable, and column-computable unitary to be
implementable with unit cost. By the results above, the computations
achievable with polynomial cost in this model can all be simulated
efficiently by quantum circuits. That is, the problems solvable in
polynomial time in this model are all contained in BQP. Conversely,
quantum gates are row-sparse, column-sparse, row-computable, and
column-computable unitaries due to their tensor product
structure. Thus the sparse unitary model is equivalent to BQP.

We next show that the quantum Turing machine model, anyonic models,
permutational quantum computation, and discrete time quantum walks all
lie within the sparse unitary model of quantum computation. Thus we
obtain as an immediate corollary to our efficient implementation for
sparse unitaries the fact that all of these models are contained in
BQP. Previously, these containments were each proved by different
methods, some fairly complicated\cite{Freedman, AJL, Yao,
  Jordan_permute, Childs_thesis}.

\subsection{Quantum Turing Machines}

Following \cite{Bernstein_Vazirani}, a quantum Turing
machine consists of a two-way infinite tape such that each site has a
finite alphabet $\Sigma$ of states, and a head that moves along the
tape one step at a time, manipulating the states of the sites and transitioning
between a finite set $Q$ of internal states. The location of the head,
the strings written on the tape, and the internal state of the head
are all allowed to go into superposition. The dynamics of the quantum
Turing machine is determined by a transition rule $\Delta: Q \times
\Sigma \times Q \times \Sigma \times \{L,R\} \to \mathbb{C}$. If $q
\in Q$ is the current state of the head, $\sigma \in \Sigma$ is
the letter on the tape at the current location of the head, and $d \in
\{L,R\}$, then $\Delta(q,\sigma,q',\sigma',d)$ is the amplitude for
the head to change the letter on the tape to $\sigma'$, transition into the
internal state $q'$, and move one step left or right on the tape
depending on the value of $d$. (In a slight variant, some definitions
also allow the head to stay in place.) $\Delta$ is not
allowed to be arbitrary. The evolution it induces on the combined
state space of the head and tape must be unitary. Furthermore, to
prevent the ``smuggling'' of uncomputable quantities into the model,
the amplitudes are required to be efficiently computable.

Let $U_\Delta$ be the unitary transition matrix induced by applying
the transition rule defined by $\Delta$ once. If we consider the
entire infinite tape, then $U_\Delta$ is
infinite-dimensional. However, after $t$ computational steps, the
number of tape locations that can have been accessed is at most
$2t+1$. Thus, the dynamics at each of the first $t$ steps is
completely captured by an $m \times m$ truncated
matrix $U_\Delta^{(t)}$ with $m = (2t+1) |Q|
|\Sigma|^{2t+1}$. In general, the matrix
$U^{(t)}_\Delta$ lacks any simple tensor product structure
or block diagonality. However, it is sparse. Specifically, by the rule
$\Delta$, a given configuration has nonzero transition amplitude to
only $2 |\Sigma| |Q|$ other configurations. (Or $3 |\Sigma| |Q|$ if
the head is allowed to stay in place.) From initial state $\ket{x}$,
the computation performed by the quantum Turing machine after $s < t$ steps is 
$\left( U^{(t)}_\Delta \right)^s \ket{x}$. To simulate
this, we can use our general technique for sparse unitaries. There is
one extra technicality: truncation can result in a matrix
$U^{(t)}_\Delta$ that is not completely unitary. As shown
in appendix \ref{trunc}, one can achieve exponentially small error by
using $U_\Delta^{(2t)}$ rather than $U_\Delta^{(t)}$ in the
Hamiltonian of equation \ref{hu}.

\subsection{Topological and Permutational Computing}

In topological quantum computation, one computates by
braiding anyons. For example, it is thought that the $SU(N)_k$
Chern-Simons anyons may arise as quasiparticle excitations in certain
quantum Hall systems, and these may provide a robust way of performing
quantum computation. Braiding $n$ anyons induces a unitary
representation of the $n$-strand braid group $B_n$. In addition to
braiding, one can also fuse and split anyons. As described in
\cite{Kauffman, Jordan_permute}, the fusion rules provide a natural
basis for the state preparations prior to braiding and the
measurements at the end of the topological computation. 

The braid group $B_n$ on $n$ strands is generated by
$\sigma_1,\ldots,\sigma_{n-1}$, where $\sigma_j$ is the clockwise 
crossing of the neighboring strands $j$ and $j+1$. For the
$SU(N)_k$ Chern-Simons anyons, in the basis defined by the anyonic
fusion rules, the representations of the generators
$\sigma_1,\ldots,\sigma_{n-1}$ are each direct sums of easily
computable $2 \times 2$ blocks. Thus, the representations of these
generators are sparse row- and column-computable unitaries, so any
braid of polynomially many crossings can be simulated efficiently in
the sparse unitary model. Conversely, any quantum circuit of
polynomially many gates can be simulated by a braid of polynomially
many crossings\cite{Freedman, Freedman2, AJL}. The quantum algorithms
for estimating Jones and HOMFLY polynomials also work by implementing
these same representations of $B_n$ using quantum circuits\cite{AJL,
  Wocjan_Yard, Jordan_Wocjan}.

Recently, a model of computation has been proposed based on the
permutation of spin-1/2 particles in states of definite total angular
momentum\cite{Jordan_permute}. This model is closely analogous to
topological quantum computation. The transformations induced by
permuting $n$ particles form a unitary representation of the symmetric
group $S_n$. The basis for state preparation and measurement come from
the fusion rules for adding angular momentum (\emph{i.e.} for the
combination of irreducible representations of $SU(2)$). $S_n$ is
generated by $s_1,\ldots,s_{n-1}$, where we imagine $S_n$
permuting $n$ objects arranged on a line, and $s_j$ exchanges the
$j\th$ and $(j+1)\th$ objects. The resulting representations of
$s_1,\ldots,s_n$ in the fusion basis has the same $2 \times
2$ block diagonal structure as in topological quantum
computing. These computations thus also fall into the sparse unitary
framework.

\subsection{Subgroup Adapted Bases}

The block diagonal structure of the representations of the generators
of $S_n$ and $B_n$ is not a coincidence. This structure stems from the
fact that the fusion bases are subgroup adapted to the chain of
subgroups $S_n \supset S_{n-1} \supset \ldots \supset S_2$ in the case
of $S_n$ and $B_n \supset B_{n-1} \supset \ldots \supset B_2$ in the
case of $B_n$. As described in \cite{generalft} such subgroup adapted
structure causes the irreducible representations of a suitable set of
generators for many groups to be direct sums of constant or polynomial
size matrices. Efficient implementations of the resulting sparse
unitaries are an essential ingredient of many non-Abelian quantum
Fourier transforms including Beals' efficient quantum Fourier
transform over the symmetric group\cite{Beals}.

\subsection{Quantum Walks}

Discrete time quantum walks are a quantum analogue to classical random
walks. For example, consider the standard discrete quantum walk on the
torus $\mathbb{Z}/n\mathbb{Z}$. The Hilbert space is spanned by
$\ket{x,i}$, where $x \in \{0,1,2,\ldots,n-1\}$ are the sites on the
torus, and $i \in \{0,1\}$ is the value of an ancilliary degree of
freedom called the coin\cite{Childs_coins, Ambainis_coins}. Each step
in the walk consists of two substeps. First, a Hadamard gate is
applied to the coin. Then, the walker is moved one step to the right
or left depending on the value of the coin:
\begin{equation}
\label{step}
\begin{array}{rcl}
\ket{x,0} & \to & \ket{x-1,0} \\
\ket{x,1} & \to & \ket{x+1,1}.
\end{array}
\end{equation}
(Here the additions and subtractions are done modulo $n$.) The above
two steps yield a $2n \times 2n$ unitary matrix with two 
nonzero entries in each row or column, and which is not
combinatorially block diagonal. More generally, one can implement a
discrete time quantum walk on any regular graph of degree $d$ by using
a coin of dimension $d$. For constant or polynomial $d$, each step of
the quantum walk is the application of a sparse unitary. This places
quantum walks within the sparse unitary framework.

\section{Conclusion}

As shown above, essentially all known discrete models of quantum computation
are special cases of our sparse unitary model. Thus, as an immediate
corollary to our result we find that all of these models are
efficiently simulatable by quantum circuits. Previously the proofs of
this for the various models were spread over many papers. Although
some of these proofs can be made by appealing to the previously known
method for implementing combinatorially block diagonal
unitaries, the proofs for discrete time quantum walks and Turing
machines cannot. Discrete time quantum walks can be efficiently
simulated on quantum circuits using elementary techniques
\cite{Ambainis_coins}. However in the case of Turing machines, our
result on sparse unitaries provides a substantial simplification over
previously known proofs. Beyond this, we hope that the sparse unitary
model of computation will be useful for the discovery of new fast
quantum algorithms.

\section{Acknowledgements}

We thank Dominik Janzing for helpful discussions.  While
preparing this manuscript we learned from Andrew Childs that he has
independently formulated the construction of equations \ref{hu} though
\ref{final} in unpublished work. S.J. gratefully acknowledges support
from the Sherman Fairchild foundation and the National Science
Foundation under grant PHY-0803371. P.W. gratefully acknowledges the
support by NSF grants CCF-0726771 and CCF-0746600.

\appendix

\section{Truncation of Turing Machines}
\label{trunc}
Corresponding to the transition rule $\Delta$ of a quantum Turing
machine, we define a truncated transition matrix $U_\Delta^{(t)}$
acting on a finite-dimensional Hilbert space. Specifically, we define
$U_\Delta^{(t)}$ by its matrix elements as follows.
\[
\begin{array}{l}
\bra{p',q',s'} U_\Delta^{(t)} \ket{p,q,s} = \vspace{5pt} \\
\Delta(q,s,q',s',L) \delta_{p',p-1} + \Delta(q,s,q',s',R)
\delta_{p',p+1} \end{array}
\]
Here $p$ and $p'$ are head positions, allowed to range from $-t$ to
$t$, $q$ and $q'$ are internal states of the head, and $s$ and $s'$
are states of a finite tape consisting of the cells
$-t,-t+1,\ldots,t$. In general, $U_\Delta^{(t)}$ is not quite unitary
because the full Turing machine has nonzero amplitudes to transition
to head positions outside $\{-t,\ldots,t\}$ and these have been cut
off. Nevertheless, the corresponding Hamiltonian
\[
H_{(t)} = \left[ \begin{array}{cc}
0 & U_\Delta^{(t)} \\
(U_\Delta^{(t)})^\dag & 0 \end{array} \right]
\]
is Hermitian and sparse. Because $U_\Delta^{(t)}$ is not
unitary, equation \ref{sincos} does not hold. If we start
with the head at site $r$ and time evolve for time $\pi/2$ with
$H_{(2t)}$, the resulting state is 
\begin{equation}
\label{taylor}
e^{-i H_{(2t)} \pi/2} \ket{r,q_0,s_0} = \sum_{n=0}^{\infty}
\frac{(-i)^n}{n!} \left( \frac{\pi}{2} \right)^n H_{(2t)}^n
\ket{r,q_0,s_0}.
\end{equation}
An application of the transition rule $\Delta$, and hence an
application of $U_\Delta^{(2t)}$ or $H_{(2t)}$,
moves the head at most one step to the left or right. Hence, for any
$r,n < t$, $H_{(2t)}^n \ket{r,q_0,s_0} = H^n
\ket{r,q_0,s_0}$. Thus in applying $e^{-i H_{(2t)} \pi/2}$ to any
state $\ket{\psi_t}$ with support only at head positions between $-t$
and $t$, the truncation error only appears at orders $t$ and higher in
the Taylor series \ref{taylor}. Thus, the truncation error
$\mathcal{E}_{\mathrm{trunc}}$ is
\begin{eqnarray*}
\mathcal{E}_{\mathrm{trunc}} & = &\| (e^{-i H_{(2t)} \pi/2} - e^{-i
  H \pi/2}  )\ket{\psi_t} \| \\
& = & \| \sum_{n=t}^\infty
  \frac{(-i)^n}{n!}(\pi/2)^n (H_{(2t)}^n - H^n) \ket{\psi_t} \| \\
 & \leq & \sum_{n=t}^\infty \left(\frac{\pi}{2}\right)^n \frac{1}{n!}
 \left( \| H_{(2t)}\|^n + \| H \|^n \right). 
\end{eqnarray*}
Applying $\| H \| = 1$ and $\| H_{(2t)} \| \leq 1$ yields
\[
\mathcal{E}_{\mathrm{trunc}} \leq 2 \sum_{n=t}^\infty
\left(\frac{\pi}{2}\right)^n \frac{1}{n!} = e^{-\Omega(t)}.
\]
To simulate the dynamics of the Turing machine for $t$ steps, one must
apply the unitary transition rule $t$ times. If the head starts at
location $0$ then the state during the first $t$ time steps always has
support only on head locations between $-t$ and $t$. Thus the above
error analysis holds at each step, and the total error is therefore at
most $t e^{-\Omega(t)}$.

\bibliography{unitaries}

\end{document}